\documentclass[aps,pre,nofootinbib,showpacs,twocolumn]{revtex4-1}%
\usepackage{graphicx}
\usepackage{color}

\begin{document}
\title{Thermally driven classical Heisenberg chain with a spatially varying magnetic field: Thermal rectification and
Negative differential thermal resistance} 
\author{Debarshee  Bagchi}
\email[E-mail address: ]{debarshee.bagchi@saha.ac.in}
\affiliation{Condensed Matter Physics Division, Saha Institute of Nuclear Physics,
Kolkata, India.}
\date{\today}

\begin{abstract}
Thermal rectification and negative differential thermal resistance are two important features
that have direct technological relevance. In this paper, we study the classical one dimensional
Heisenberg model, thermally driven by heat baths attached at the two ends of the system, and in
presence of an external magnetic field that varies monotonically in space. Heat conduction in this
system is studied using a local energy conserving dynamics. It is found that, by suitably tuning
the spatially varying magnetic field, the homogeneous symmetric system exhibits both thermal
rectification and negative differential thermal resistance. Thermal rectification, in some parameter
ranges, shows interesting dependences on the average temperature $T$ and the system size $N$ - rectification
improves as $T$ and $N$ is increased. Using the microscopic dynamics of the spins we present a
physical picture to explain the features observed in rectification as exhibited by this system
and provide supporting numerical evidences. Emergence of NDTR in this system can be controlled
by tuning the external magnetic field alone which can have possible applications in the fabrication
of thermal devices.
\end{abstract}

\maketitle

\section{Introduction}
\label{Intro}

Thermal rectification (TR) \cite{phononics,control} is an important property that has been extensively studied in a
variety of nonlinear systems \cite{diode1, diode2, morse, FK-FK, FK-FPU, graded1, graded2, 2D, 3D,rect_review} in
recent times. A thermally driven system can be so designed such that the thermal current through the system has unequal
values when direction of thermal bias is reversed - the heat conduction is asymmetric. Thus the system behaves
as a good heat conductor in one direction and a good insulator in the opposite direction. Thermal rectification
owes its origin to the nonlinearity of the system and to its spatial asymmetry. Analogous to its electrical
counterpart, the thermal rectifier is considered to be a crucial building block and therefore has an important
role to play in the fabrication of sophisticated thermal devices. 

Negative differential thermal resistance (NDTR) \cite{phononics,control} is a counter intuitive phenomenon, predicted
in the heat conduction studies, where the steady state thermal current decreases as the temperature difference
across a system is increased. In the recent decades a lot of attention has been devoted to study NDTR in nonlinear
lattices. However, in spite of enormous efforts the underlying physical mechanism that generates NDTR in nonlinear
system is still not satisfactorily understood. A lot of mechanisms have been proposed and many unresolved
questions regarding the emergence of NDTR, such as the mismatch of the phonon bands \cite{transistor,gate},
role of interface \cite{origin}, ballistic-diffusive transport \cite{ballistic_NO, ballistic_yes}, role of
momentum conservation \cite{anomalous} presence of a critical system size and a  transition from the
exhibition to the nonexhibition of NDTR \cite{transition} and scaling \cite{scaling} are still being
explored. NDTR is considered to be of immense technological importance in the working of recently
proposed thermal devices such as the thermal transistors \cite{transistor}, thermal logic gates \cite{gate},
thermal memory elements \cite{memory} etc. 
Theoretical studies using mostly numerical simulation have been employed extensively to study both these feature
in many nonlinear lattice systems, few examples are the Frenkel-Kontorova model \cite{FK-FK, transition}, the
$\phi^4$ model \cite{origin}, the Fermi-Pasta-Ulam chain \cite{FK-FPU, anomalous}, the Morse lattice \cite{morse}.

Linear systems, such as the coupled harmonic oscillators, do not exhibit TR or NDTR. Surprisingly it has
been recently found that linear graded systems, such as a harmonic oscillator chain with linearly increasing masses,
show both TR and NDTR \cite{graded1, graded2}. In fact, gradual mass-loaded carbon and boron nitride nanotubes
have already been used effectively to fabricate a thermal rectifier \cite{diode}. These functional graded materials
have been considered to be of huge technological relevance since these materials can be purposely manufactured and
have many intriguing optical, electrical, mechanical and thermal properties \cite{graded1}.

Motivated by the concept of these functional graded materials, in this paper, we study thermal transport in the 
classical Heisenberg model \cite{fisher,joyce} coupled to heat baths and in presence of a spatially varying
magnetic field, and investigate TR and NDTR. Both of these features have been shown to emerge in the Heisenberg
spin chain previously \cite{NDTR} but by a different approach; in this paper we present a new route to obtain these
features.

Investigation of TR and NDTR in spin systems have been carried out only in a very few
other works such as the two dimensional classical Ising model \cite{Ising2D} and quantum
spin systems \cite{IsingQnt} very recently. These systems are quite simple and although they are
helpful in understanding the underlying physical mechanism, nevertheless these are not very realistic; 
the Heisenberg model is a comparatively more realistic spin model for a magnetic insulator.
Also, in both cases the system under consideration consists of two dissimilar segments coupled to
each other. This scheme requires one to carefully fabricate the junction as it has been argued that
TR and NDTR are crucially dependent on the junction properties \cite{ballistic_yes, transition}
which is difficult to implement this in real systems. It was also initially believed that
NDTR can not be obtained in a symmetric system \cite{symm_NO}. However later studies clearly showed
that NDTR can be obtained from systems even without structural inhomogeneity \cite{scaling,
symm_YES_CNT}.

The advantage of our proposal is that it is much simpler to implement, easy to manipulate over a wide range,
and should also be realizable in practice. Firstly, one does not need to specially design the system,
unlike the case of two segment nonlinear lattices (with an interface) or graded systems mentioned above - 
one has to fabricate such a system with precise specifications which might be technologically more challenging
and restrictive in applicability.
Secondly, using spin systems one can control TR and NDTR over a wide range by tuning only an external magnetic
field and so, in contrast to previous works, no special engineering of the system is required in our case. TR
in this system also shows interesting anomalous dependences, as we shall discuss, that can be of technological
relevance.

The organization of the rest of the paper is as follows. In the next section Sec \ref{model} we describe
our model and the numerical scheme employed to study the system. Thereafter we present our results in
Sec. \ref{results}. Finally, in Sec \ref{conclusion} we conclude with a brief summary of our results and a
discussion.

\section{Model and numerical scheme}
\label{model}
Consider classical Heisenberg spins $\{\vec{S_i}\}$ (three-dimensional unit vectors) on a one-dimensional
lattice of length $N$ $(1 \le i \le N)$ with nearest neighbor interaction. The Hamiltonian of the
system is
\begin{equation}
\mathcal{H} = -K \sum_{i=1}^{N-1} \vec S_i\cdot \vec{ S}_{i+1} - \sum_{i=1}^{N}\vec h_i \cdot \vec S_i
\label{ham}
\end{equation}
where the spin-spin interaction are taken to interact ferromagnetically $K > 0$ (we have set $K$ to unity
for our results without any loss of generality). The second term in Eq. (\ref{ham}) is due to a spatially
varying magnetic field $\vec h_i$ that acts on all the spins. The equation for the time evolution of the
spin vectors can be written as
\begin{equation}
\frac d{dt} {\vec{ S}_i} = \vec{S}_i \times \vec{B}_i
\label{eom}
\end{equation}
where $\vec{B}_i = \vec{S}_{i-1} + \vec{S}_{i+1} + \vec h_i$  (with $K = 1$) is the local molecular field
experienced by $i$-th spin vector.

To drive the system out of equilibrium we couple the ends of the system to two heat baths.
This is implemented by introducing to additional spins at sites $i = 0$ and $i = N+1$. The
bonds between ($\vec{S}_0, \vec{S}_1$) and ($\vec{S}_N, \vec{S}_{N+1}$) at two extreme ends
of the system behave as stochastic thermal baths. The left and right thermal baths are in
equilibrium at their respective temperatures, $T_l$ and $T_r$ and the bond energies
$E_l = -\vec S_0 \cdot  \vec S_1$ and $ E_r = -\vec S_N \cdot  \vec S_{N+1}$ have Boltzmann
distribution $P(E) \sim \exp(-E/T)$.
The average energies of the two extreme bonds read $\langle E_{l}\rangle=-\mathcal{L}(T_l^{-1})$
and $\langle E_{r}\rangle=-\mathcal{L}(T_r^{-1})$, $\mathcal{L}(x) = \coth (x) - 1/x$ being the
standard Langevin function.

We investigate the steady state transport properties of the Heisenberg model by numerically computing
the steady state thermal current using the discrete time odd even (DTOE) dynamics \cite{our, NDTR}.
The DTOE dynamics alternately updates the spins belonging to the odd and even sites of the lattice
using a spin precession dynamics given by
\begin{equation}
\vec{S}_{i,t+1} = \left[\vec{S} \cos \phi + (\vec{S} \times \hat{B}) \sin \phi + (\vec{S}\cdotp\hat{B})
\hat{B}(1-\cos \phi) \right]_{i,t}
\label{precess}
\end{equation}
where $\hat{B}_i = \vec{B}_i/|\vec{B}_i|$, $\phi_i = |\vec{B}_i| \Delta t$ and $\Delta t$ is the
time-step increment \cite{our}.

Numerically, the leftmost spin $\vec S_0$ is updated with the even spins and the rightmost
spin $\vec S_{N+1}$ is updated with the odd (even) spins for even (odd) $N$. The bond energy
between $\vec{S}_0$ and $\vec{S}_1$ is refreshed from a Boltzmann distribution and thereafter
the spin $\vec{S}_0$ is reconstructed using the relation $E_0 = - \vec S_0 \cdot \vec S_1$.
Note that, during this update $\vec{S}_1$ is not modified (as it belongs to the odd sublattice).
Similarly the other end is also updated. This sets the temperatures of the two ends of the lattice
to our desired values. A thorough discussion of the DTOE scheme and numerical implementation of
the thermal baths can be found in Ref. \cite{our}.

The computation of the steady state thermal current is done as described in the following.
The energy of the $i$-th bond $E^o_i$ measured after the odd spin update is not equal to 
$E^e_i$ measured after the subsequent even spin update, where
$E_i = - \vec S_i \cdot \left[\vec S_{i+1} + \vec h_i\right]$ is the energy density.
This difference $(E^e_i - E^o_i)$ is the measure of the energy crossing the $i$-th bond in
time $\Delta t$ (we set $\Delta t$ to unity \cite{our}). The steady state thermal current
$j$ (rate of flow of energy) is site independent and is computed in this scheme \cite{our,NDTR}
using 
\begin{equation}
j = \langle E^e_i - E^o_i \rangle.
\label{J}
\end{equation}
Note that Eq. (\ref{J}) is consistent with the definition of current obtained from the continuity equation
\cite{our}. We define a total current $J = jN$ and all the results obtained are presented below in terms of
this total current.

\section{Results}
\label{results}

\subsection{Thermal Rectification}
\begin{figure}[h]
\hskip-0.33cm
\includegraphics[width=3.2cm,angle=-90]{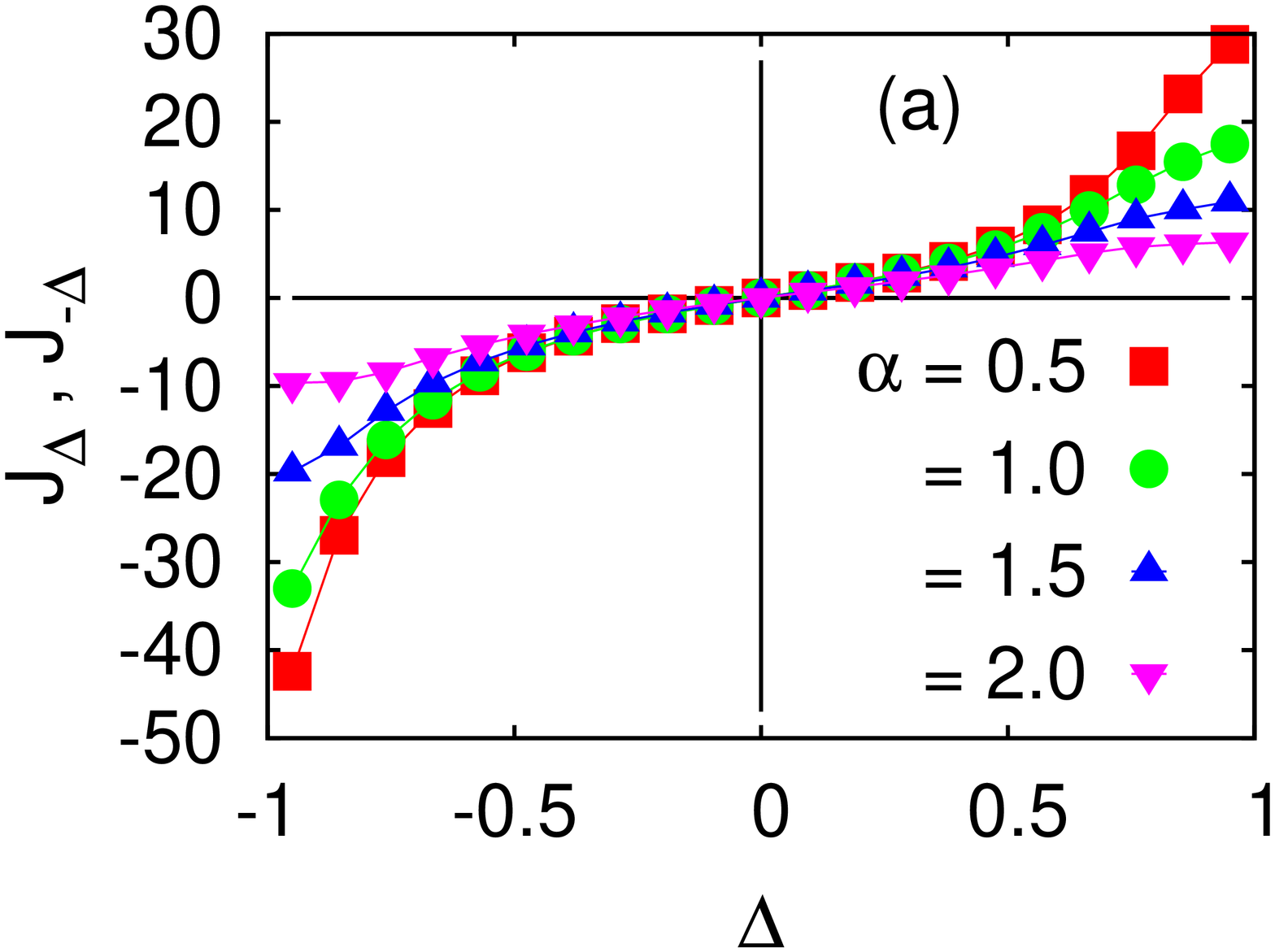}\hskip-0.175cm
\includegraphics[width=3.2cm,angle=-90]{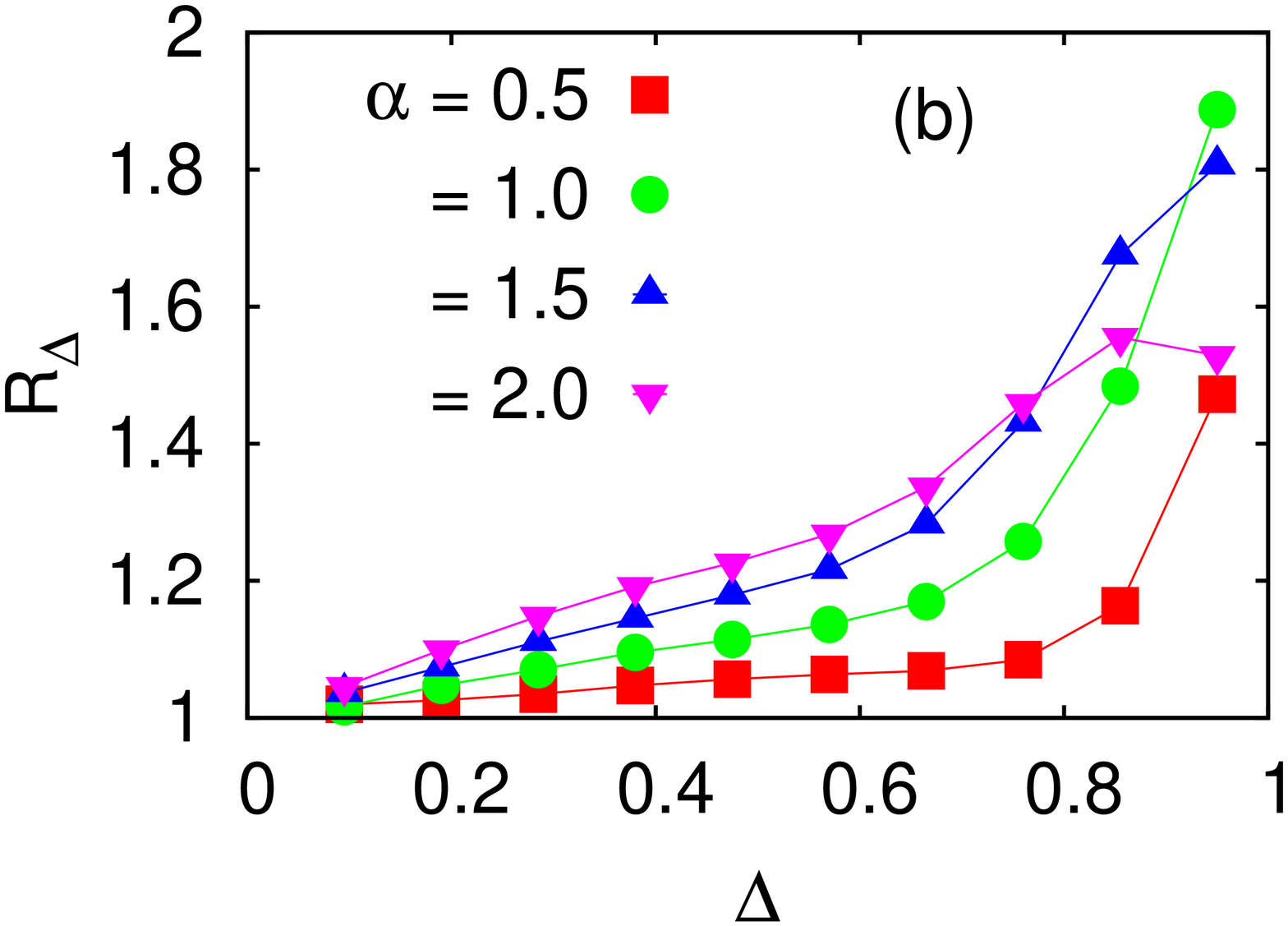}
\caption{(Color online) Variation of the total thermal current $J$ with $\Delta$ in the range $-1 < \Delta < 1$
for different values of the parameter $\alpha$. The bath temperatures are $T_l = T(1 + \Delta)$ and
$T_r = T(1 - \Delta)$ and the magnetic field $\vec h_i \equiv (0,0,h^z_i)$ varies linearly
$h^z_i = h_0 + \alpha ~ i/N$ in space. The parameters used are $T = 1$ and system size $N = 500$.}
\label{fig:rect}
\end{figure}

The temperature of the two thermal baths are set as $T_l = T(1 + \Delta)$ and $T_r = T(1 - \Delta)$, thus
the average temperature of the system is $\frac 12(T_l + T_r) = T$. The spatially varying magnetic field
$\vec h_i$ is chosen as $(0,0,h^z_i)$ and $h^z_i = h_0 + \alpha ~ i/N$ is a linearly varying field where
$1 \le i \le N$; we set $h_0=1$ for all our results in this section. 
Starting from a random initial configuration of spins we let the system evolve using the DTOE dynamics until
a steady state is reached and thereafter compute the thermal current using Eq. (\ref{J}). We consider the system
to be in {\it forward} bias for $\Delta > 0$ and in {\it backward} bias for $\Delta < 0$. 
%
The thermal current under the forward bias $J_{\Delta}$ and that in the backward bias $J_{-\Delta}$ are different in
magnitude as can be seen from  Fig. \ref{fig:rect}a. Note that the system is perfectly symmetric and homogeneous. The
asymmetric heat conduction is completely brought about by the spatially varying magnetic field. We define the
rectification ratio as $R_{\Delta} = |J_{- \Delta}/J_{\Delta}|$ which measure the of the amount of TR achieved. Thus
for poor rectification $R_{\Delta}$ is close to unity and for good rectification $R_{\Delta}$ is very large (small) if
$J_{-\Delta} \gg J_{\Delta}$ ($J_{-\Delta} \ll J_{\Delta}$). From Fig. \ref{fig:rect}b, as expected $R_{\Delta}$ is
found to increase as $\alpha$ is increased i.e., when the magnetic field varies more sharply across the system (apart
from some discrepancies for large $\Delta$).
Thus heat conduction is asymmetric i.e. $J_{\Delta} \ne J_{-\Delta}$ and the system exhibits TR.

\begin{figure}[h]
\hskip-0.3cm
\includegraphics[width=3.15cm,angle=-90]{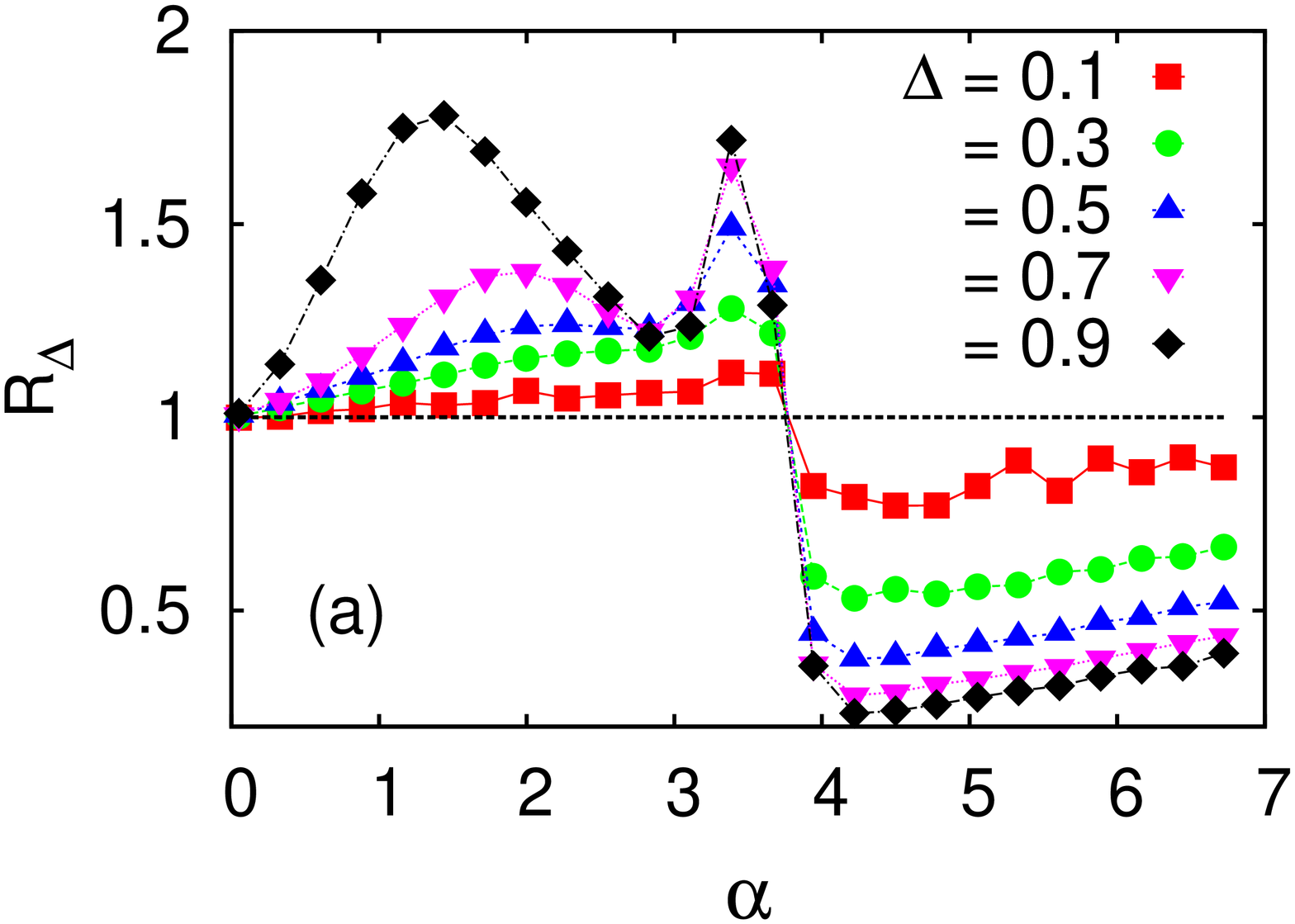}\hskip-0.15cm
\includegraphics[width=3.15cm,angle=-90]{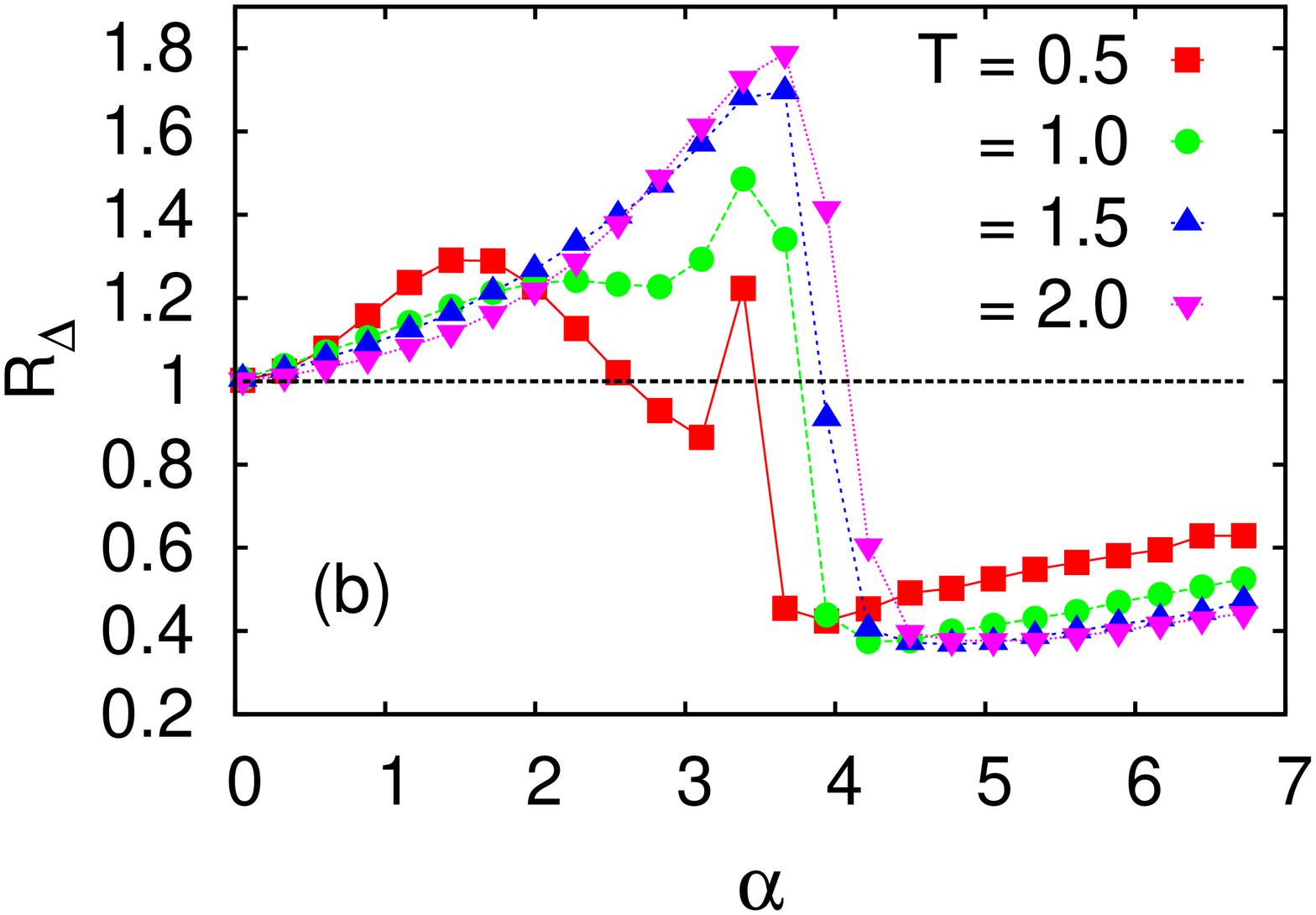}\\
\includegraphics[width=3.15cm,angle=-90]{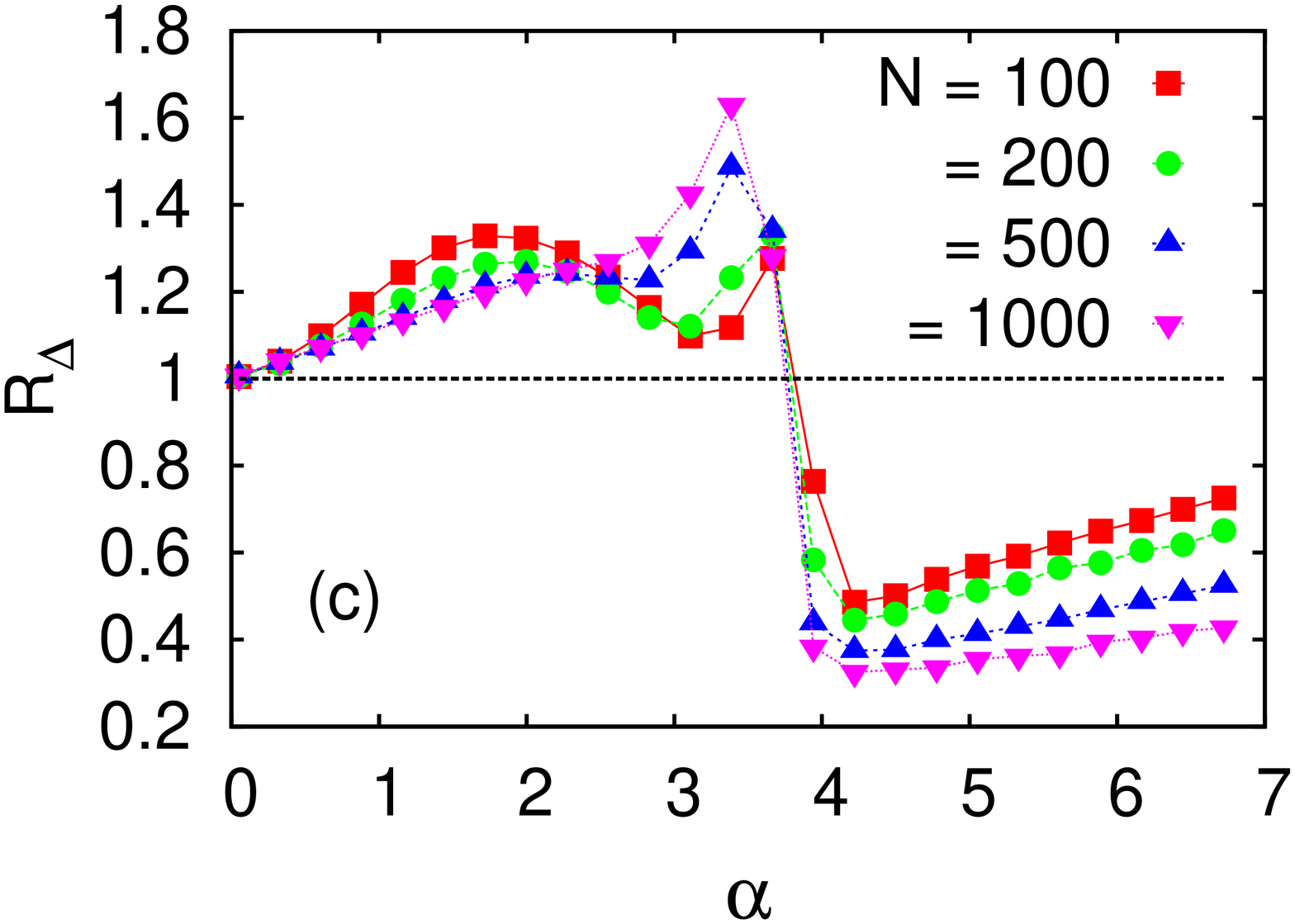}
\caption{(Color online) Variation of the rectification ratio $R_{\Delta}$ with $\alpha$ for different values of
(a) thermal bias $\Delta$ for $T = 1$ and $N = 500$; (b) temperature $T$ with $\Delta = 0.5$
and $N = 500$; (c) system size $N$ with $T = 1$, $\Delta = 0.5$. For all the cases $h_0 = 1$.}
\label{fig:RTL}
\end{figure}

An interesting feature of TR in this system is the variation of the rectification
ratio $R_{\Delta} = |J_{-\Delta}/J_{\Delta}|$ with the parameter $\alpha$. The rectification
ratio does not increase indefinitely as $\alpha$ is increased but rather shows an intriguing
nonmonotonic $\alpha$ dependence. For different values of the thermal bias $\Delta$ we compute
$R_{\Delta}$ as a function of $\alpha$ and is shown in Fig. \ref{fig:RTL}a.
For $\alpha$ in the range $0 < \alpha \lesssim \alpha_0$, we find $R_{\Delta} > 1$ initially whereas for
$\alpha \gtrsim \alpha_0$, $R_{\Delta} < 1$, where $\alpha_0$ lies roughly in the range $3.5< \alpha <4.0$
(Fig. \ref{fig:RTL}a).
For small $\Delta$, $R_{\Delta}$ increases roughly linearly for $\alpha < \alpha_0$, then drops abruptly
below $R_{\Delta}=1$ and then increases linearly towards unity again. For larger $\Delta$, $R_{\Delta}$ has an
even more complicated nonmonotonic $\alpha$ dependence but jumps from $R_{\Delta}>1$ to $R_{\Delta}<1$ at the same
$\alpha = \alpha_0$.

We look into the temperature dependence of $R_{\Delta}$ which is also very unusual. Generally rectification
is found to deteriorate as the average temperature of the system is increased \cite{phononics,control,NDTR}.
However in our case TR for higher temperature in certain range(s) of $\alpha$ is actually higher than that
for lower temperature as can be seen in Fig. \ref{fig:RTL}b. Also note that $\alpha_0$ shifts to higher $\alpha$
values as the average temperature is increased. 
Similar nontrivial dependence is seen when one studies the variation of $R_{\Delta}$ with the system size $N$.
In some $\alpha$ regime, $R_{\Delta}$ decreases as $N$ is increased whereas in some other regime we get an
{\it anomalous} size dependence as can be seen from Fig. \ref{fig:RTL}c. Thus, depending on which $\alpha$
range one is in, the $T$ and $N$ dependences can be normal ($R_{\Delta}$ approaching unity as $T$ and $L$
increases) or anomalous. This seems to be due to a complicated interplay of the imposed thermal bias and
the spatially varying magnetic field.
\begin{figure}[h]
\centering
\includegraphics[width=7.0cm,angle=0]{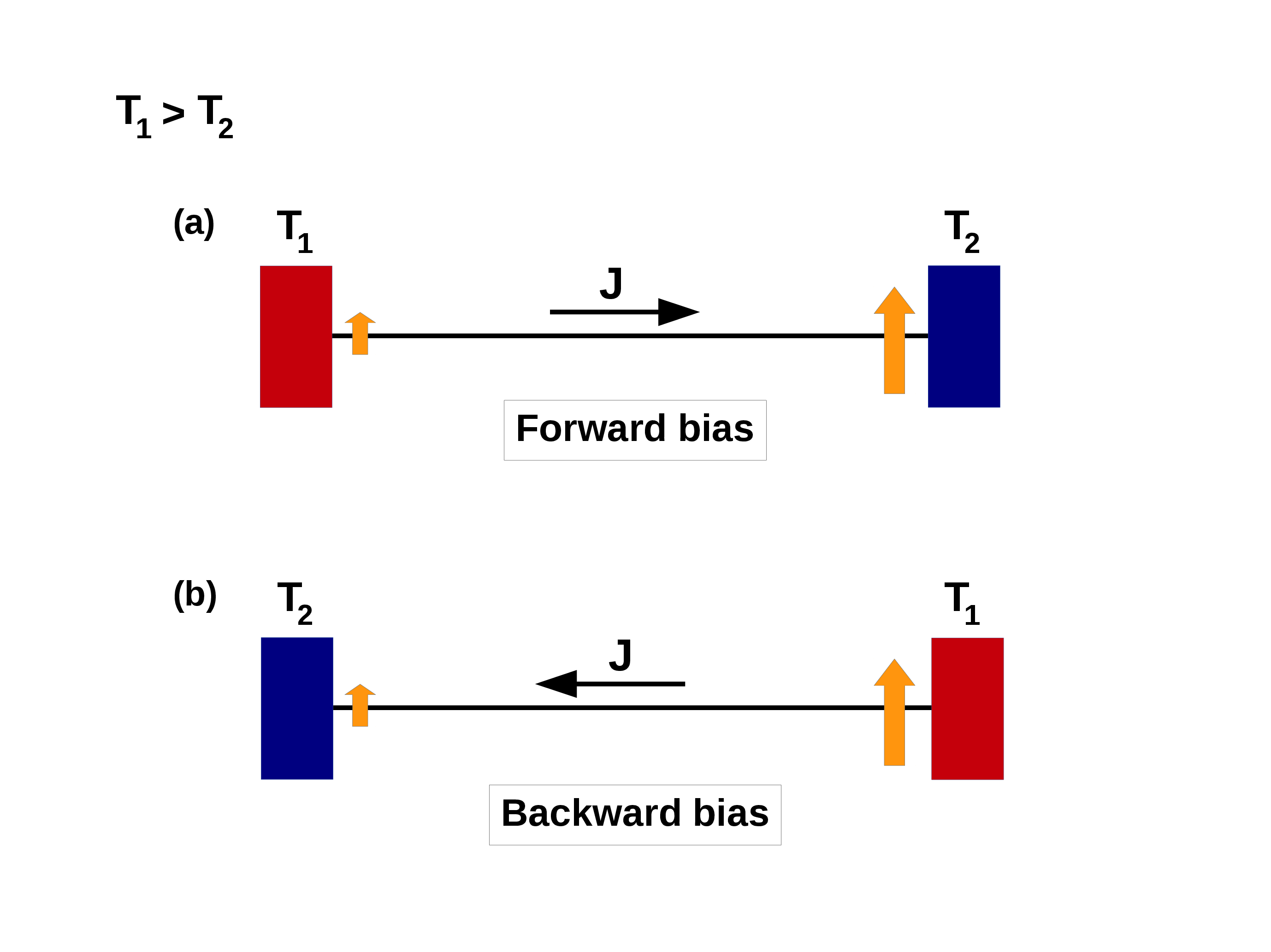}
\caption{(Color online) Schematic digram of the system in (a) forward and (b) backward bias conditions according to our
definition. The horizontal line represents the spin chain attached between the two baths with bath temperatures $T_1$
and $T_2$ where $T_1 > T_2$. The vertical arrows represent the spatially varying magnetic field which grows monotonically
as one moves from the left end of the system towards the right end.}
\label{fig:pic}
\end{figure}

\begin{figure}[h]
\hskip-0.4cm
\includegraphics[width=3.2cm,angle=-90]{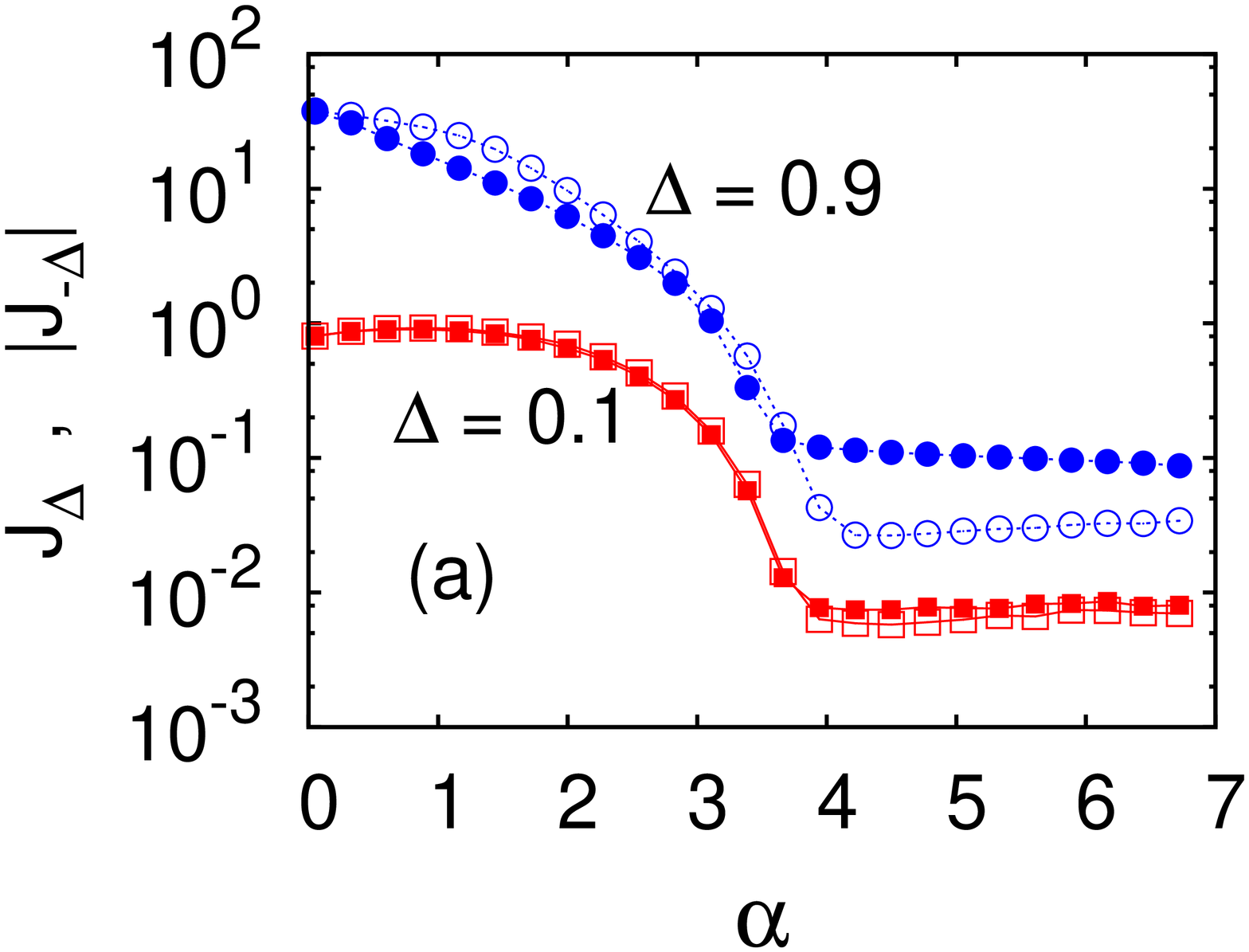}\hskip-0.4cm
\includegraphics[width=3.2cm,angle=-90]{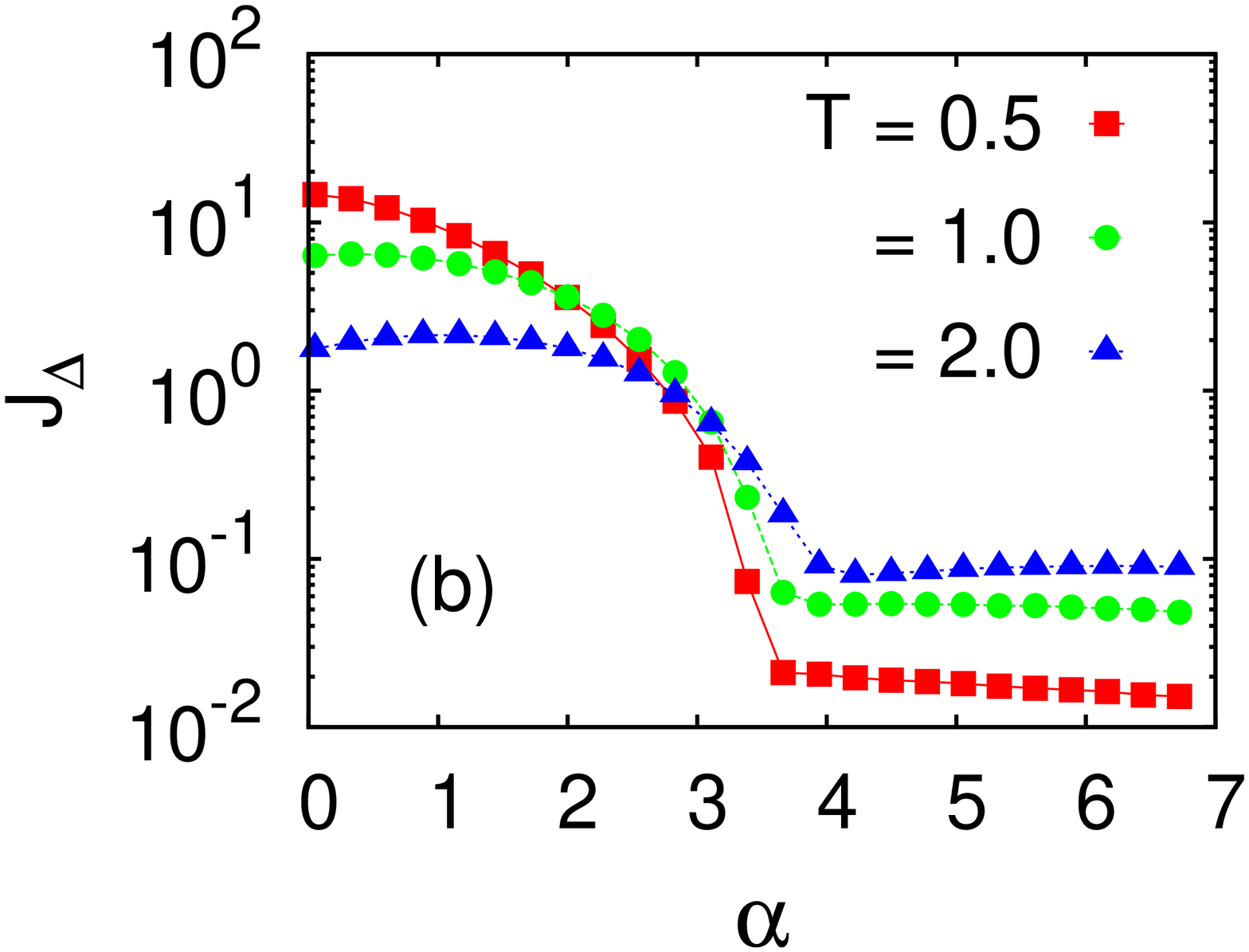}\\
\includegraphics[width=3.2cm,angle=-90]{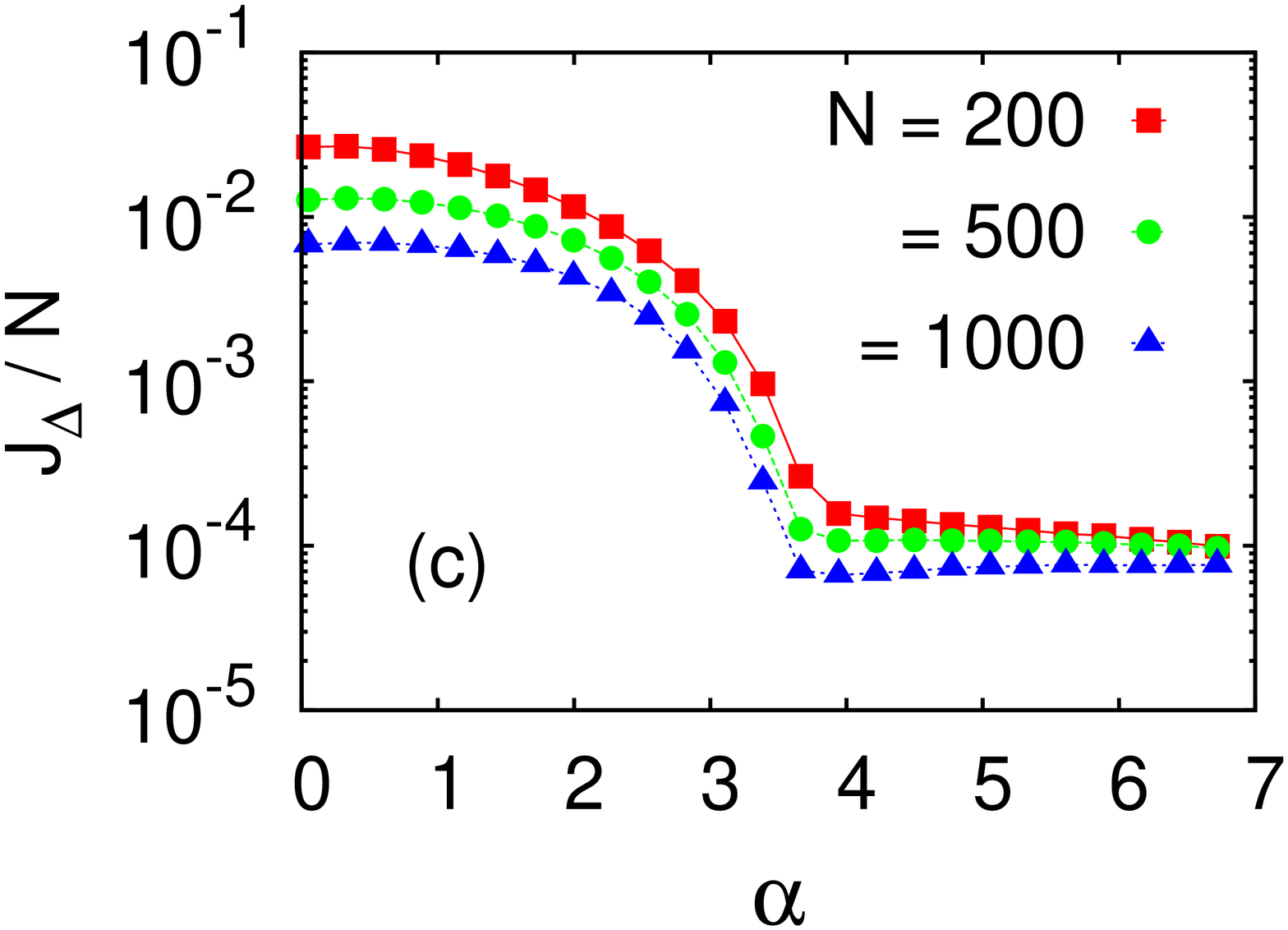}
\caption{(Color online) (a) Variation of the individual currents $J_{\Delta}$ and $|J_{-\Delta}|$ with $\alpha$ for two values of the
thermal gradient $\Delta = 0.1$ and $0.9$. Average temperatures $T =1.0$.
(b) Variation of $J_{\Delta}$ with $\alpha$ for different average temperatures $T = 0.5,1.0$ and $2.0$. Here $\Delta = 0.5$
and $N=500$ in both the cases.
(c) Variation of $J_{\Delta}$ (scaled by $N$) with $\alpha$ for different system size $N = 200, 500$ and $1000$. Here $T=1.0$
and $\Delta=0.5$. For (b) and (c) $|J_{-\Delta}|$ has similar variation as in (a).}
\label{fig:R-JTL}
\end{figure}
To understand this we look into the individual currents $J_{\Delta}$ and $J_{-\Delta}$. For relatively small
values of the magnetic field the current $J$ is higher when it flows from a higher magnetic field region to
a lower magnetic field region. This is due to the fact that the magnetic field tries to restrict the motion
of the spins and thereby inhibits the flow of energy through the system. Now according to our definition, in
the forward bias (Fig. \ref{fig:pic}a) the magnetic field increases as one approaches the colder bath - thus
the motion of the spins nearer to the right end of the system is doubly restricted - one because of the low
temperature and the other due to the higher magnetic fields. For the backward bias (Fig. \ref{fig:pic}b) however
the effect of the higher magnetic field is somewhat compensated by the hotter bath and the spins are relatively
more free to rotate in this case and therefore the system has a higher current.
Since in the steady state the current through the system is a site independent constant (a consequence of the
equation of continuity) the overall current of the system is dictated by the current carrying capacity of the
{\it weakest} bond (corresponding to the most restricted spin) and therefore the current in the forward bias is
lower than that in the backward bias. This explain why $|J_{-\Delta}| > J_{\Delta}$ for $\alpha < \alpha_0$,
as can also be seen in Fig. \ref{fig:R-JTL}a, and the rectification ratio $R_{\Delta} > 1$; $R_{\Delta}$ increases
in this region as $\alpha$ is increased because of increased asymmetry of the system.
As the magnetic field increases the current starts to decrease since the orientational {\it stiffness} of the spins
increases which restricts energy passage through the system. As the magnetic field becomes high the system goes into
a magnetic field dominated regime which limits the current carrying capacity of the system - the weaker current
$J_{\Delta}$ attains a saturation first while the relatively stronger $J_{-\Delta}$ still continues to decrease but
eventually it too attains a saturation (Fig. \ref{fig:R-JTL}a) (note that, in Fig. \ref{fig:R-JTL} the y-axis is a
logarithmic scale in all the figures).

For a lower temperature the spins of the system are more orientationally stiff and thus this domination of the magnetic
field commences at a lower value of $\alpha$. The current saturates at lower $\alpha$ (Fig. \ref{fig:R-JTL}b) and this
explains the decrease of $\alpha_0$ as the average temperature is decreased in Fig. \ref{fig:RTL}b.
With regards to the value of $\alpha_0$ there is no appreciable variation as the system size $N$ is altered as can be
seen from Fig. \ref{fig:R-JTL}c and also previously in Fig. \ref{fig:RTL}c. The system of smaller size is closer to the
ballistic limit and carries slightly more current than a system of larger size \cite{our}.

The system approaches a diffusive transport regime as the temperature $T$ or the size $N$ is increases \cite{our}, but
in the forward bias condition the approach is obviously slower than in the forward bias. This is the reason that one
obtains an improvement of rectification ($R_{\Delta}$ moves away from unity as in Fig. \ref{fig:RTL}b,c) as $T$ or $N$
is increased in the $\alpha > \alpha_0$ regime. Also, as we shall show in the following, it is the motion of the $N$th
spin that decides the value of $\alpha_0$ which therefore is independent of the length of the system to which it belongs.
This is why $\alpha_0$ remains essentially unchanged as the system size $N$ is altered and changes only when the average
temperature $T$ is changed.

Thus to summarize, there are two regimes corresponding to the two terms in the Hamiltonian (Eq. \ref{ham}): (a) a spin-spin
interaction dominated regime (or in other words, a temperature dominated regime) in the parameter range $0 < \alpha < \alpha_0$ in which
the current steadily decrease as $\alpha$ increases, and (b) a magnetic field dominated regime for $\alpha > \alpha_0$ in
which the spins have a restricted motion and the current through the system changes very negligibly as $\alpha$ is varied.
It is the complicated interplay of these two mutually opposing factors that gives rise to the interesting features that are
observed in our system.
\begin{figure}[h]
\hskip-0.3cm
\includegraphics[width=3.3cm,angle=-90]{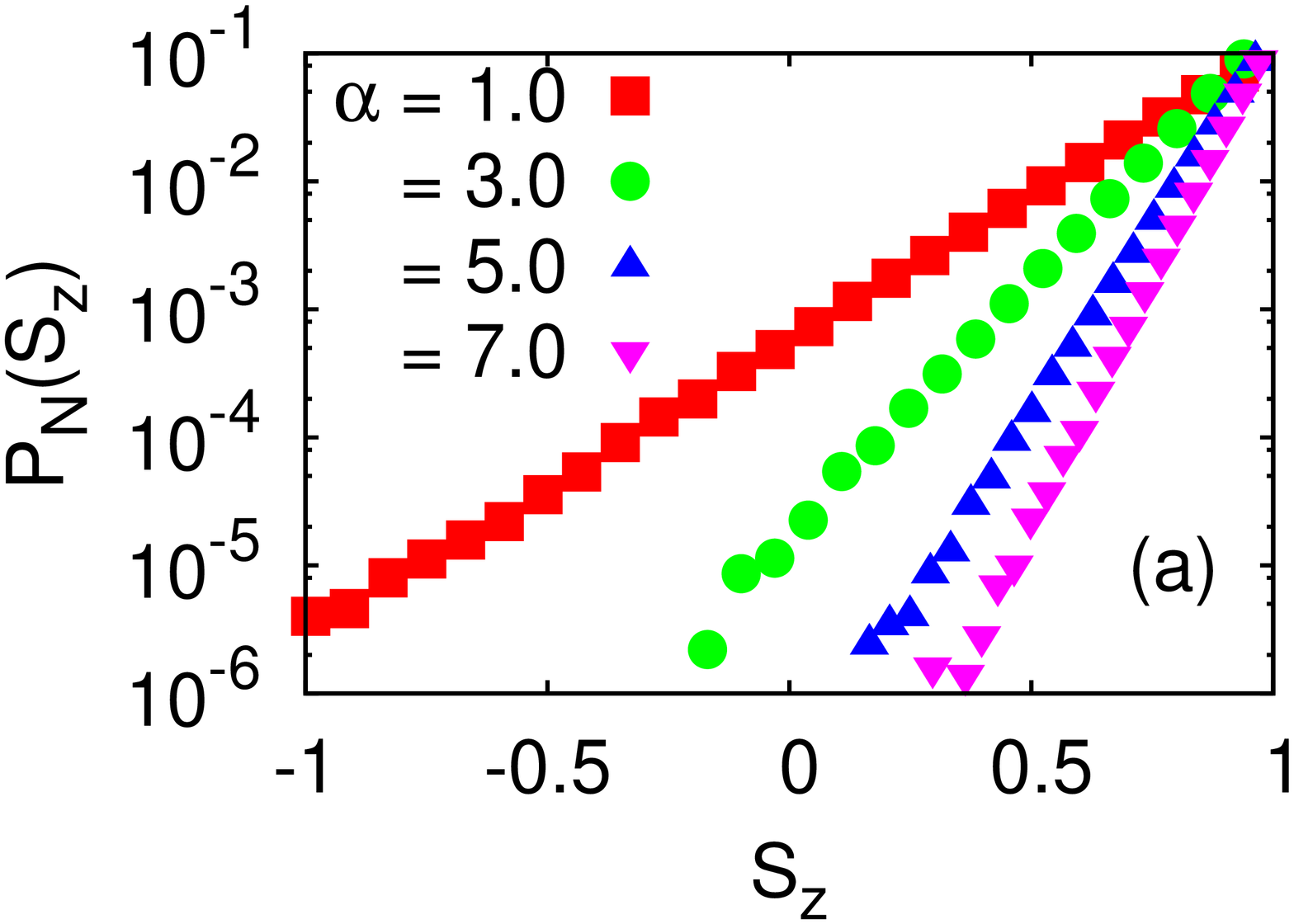}\hskip-0.3cm
\includegraphics[width=3.0cm,angle=-90]{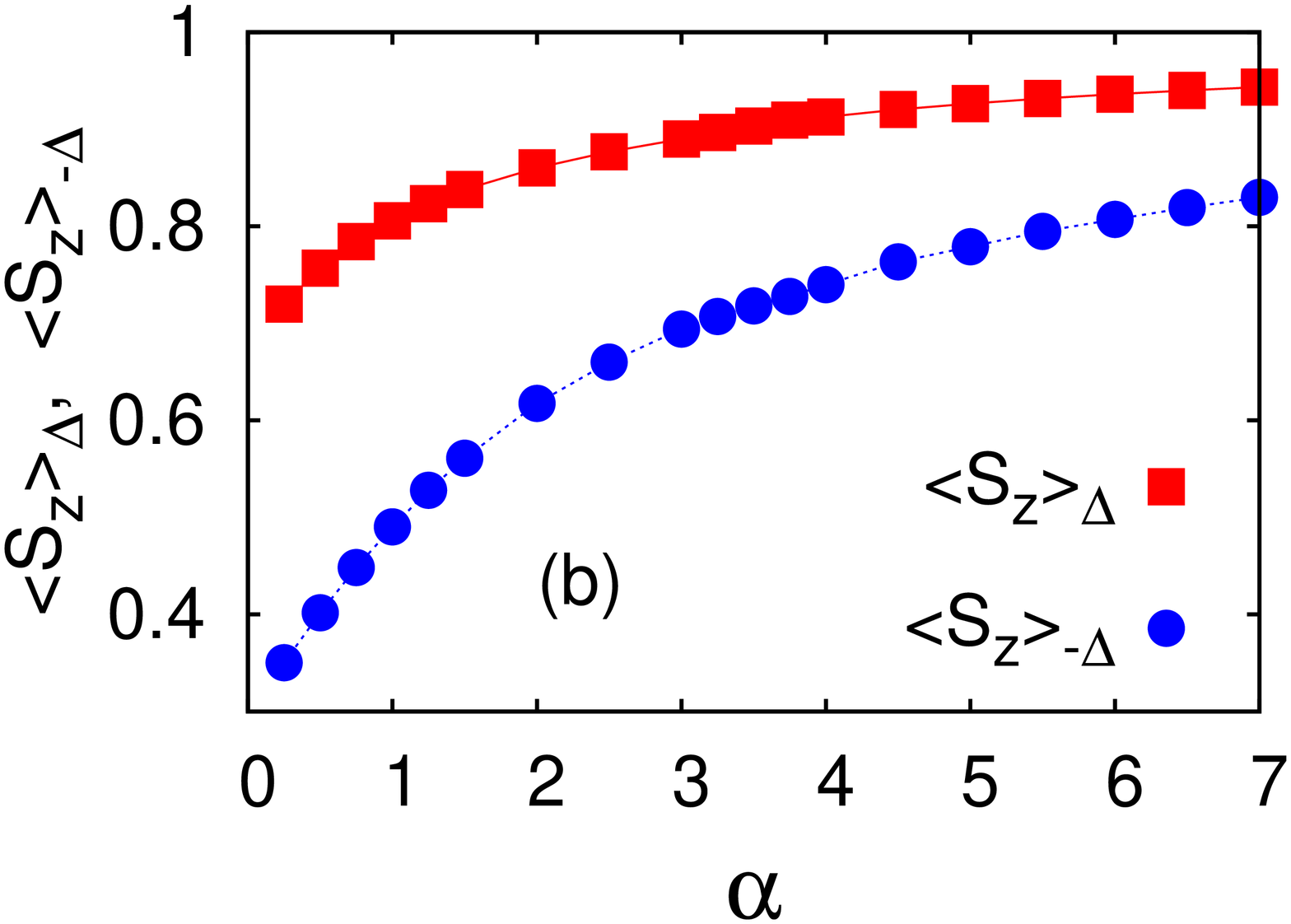}\\
\includegraphics[width=3.1cm,angle=-90]{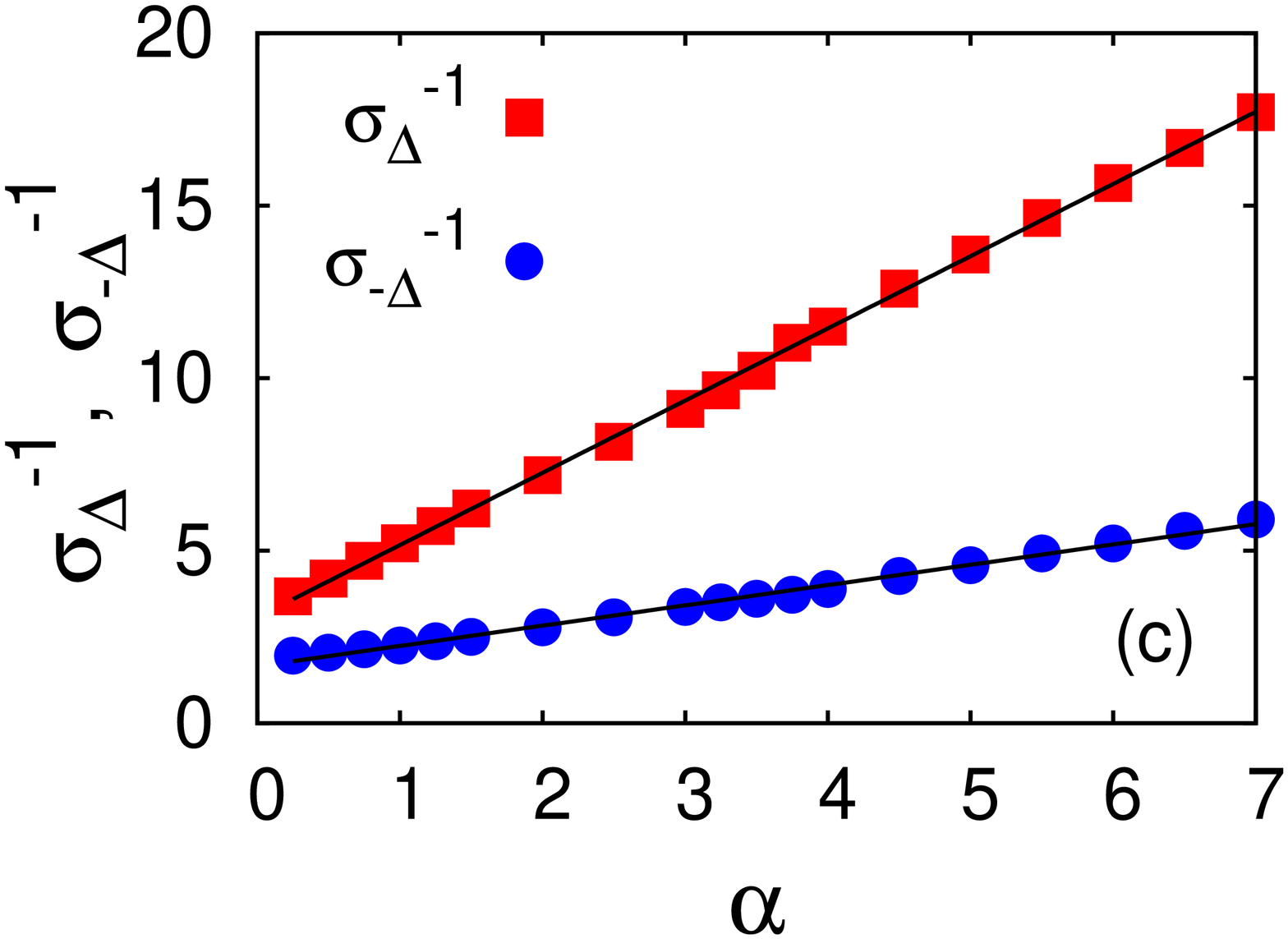}\hskip-0.3cm
\includegraphics[width=3.1cm,angle=-90]{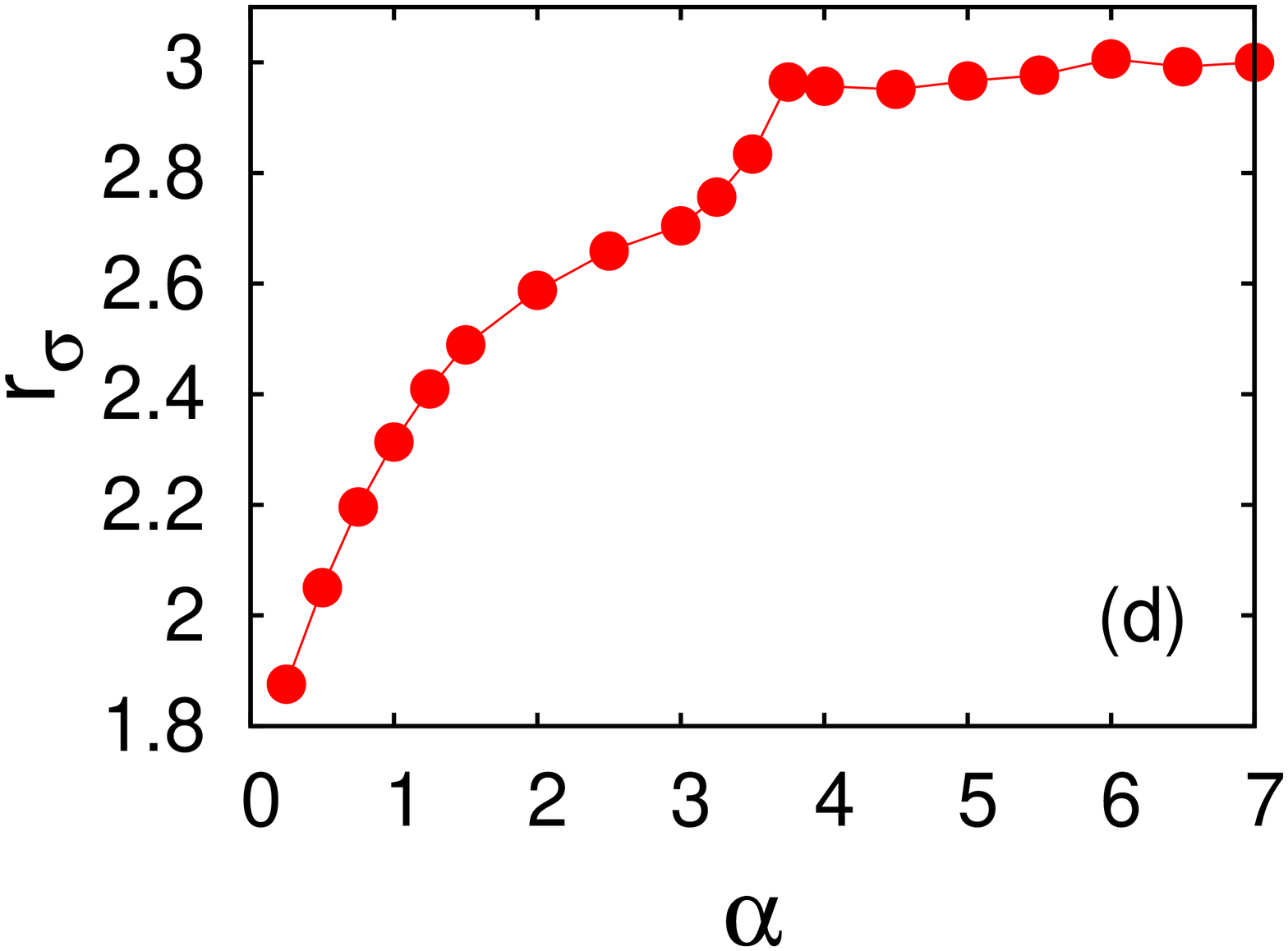}
\caption{(Color online) (a) Semi-log plot of the distribution $P_N(S_z)$ for different values of $\alpha$. 
(b) and (c) are the variation of standard deviation of the distribution $\sigma$ with $\alpha$ for the $N$th spin
corresponding to the forward and the backward bias respectively. In (d) we show $r_{\sigma} = \sigma_{-\Delta}/
\sigma_{\Delta}$ with $\alpha$. For all the cases $T=1.0$, $\Delta=0.5$ and $N = 500$.}
\label{fig:sd}
\end{figure}

In order to validate the picture just presented 
we look into the microscopic dynamics of the individual spins now. Since the most restricted
spin dictates the current (as discussed above) we look into the $S_z$ component of the $N$th spin (since the $N$th spin
experiences the largest magnetic field amongst all the spins) and show its distribution $P_N(S_z)$ as a function of the
parameter $\alpha$ in Fig. \ref{fig:sd}a. Since $S_z \equiv \cos \theta$ ($\theta$ is the polar angle and the spins being
of unit magnitude), $S_z$ lies in the range $[-1,1]$. For a spin which is completely free to orient itself in all possible
directions, the distribution would be uniform in the range $[-1,1]$. However in presence of a finite temperature and
an external magnetic field $P_N(S_z)$ has an exponential form. Note that, as $\alpha$ increases, the slope of the distribution
$\frac{d}{dS_z}\ln P_N(S_z)$ increases which signifies that the spin motion gets more and more restricted.
From the distribution we compute the average $\langle S_z \rangle$ and the standard deviation $\sigma = (\langle S_z^2 \rangle
- \langle S_z \rangle^2)^{-1/2}$ for the $N$th spin.  These two quantities are shown in Fig. \ref{fig:sd}b and Fig. \ref{fig:sd}c
respectively. The average $\langle S_z \rangle$ approaches unity  $\alpha$ increases for both the forward and backward bias
conditions, and $\langle S_z \rangle_{\Delta} > \langle S_z \rangle_{-\Delta}$.
The standard deviation $\sigma$ is an indicator of how freely a spin can rotate about the magnetic field. Thus larger the
$\sigma$ is the more is the current $J$ that the spin allows to pass through. Fig. \ref{fig:sd}c shows that $\sigma$ for
forward and backward bias decreases as $\alpha$ increases although not always monotonically for all parameters. (For the
chosen parameters for the figure, $\sigma$ approximately fits to the form $\sigma^{-1} = m \alpha + c$, where $m$ and $c$
are constants - both $m$ and $c$ values are higher for $J_{\Delta}$ than $J_{-\Delta}$). 

\vskip-0.15cm
\begin{figure}[h]
\hskip-0.3cm
\includegraphics[width=3.1cm,angle=-90]{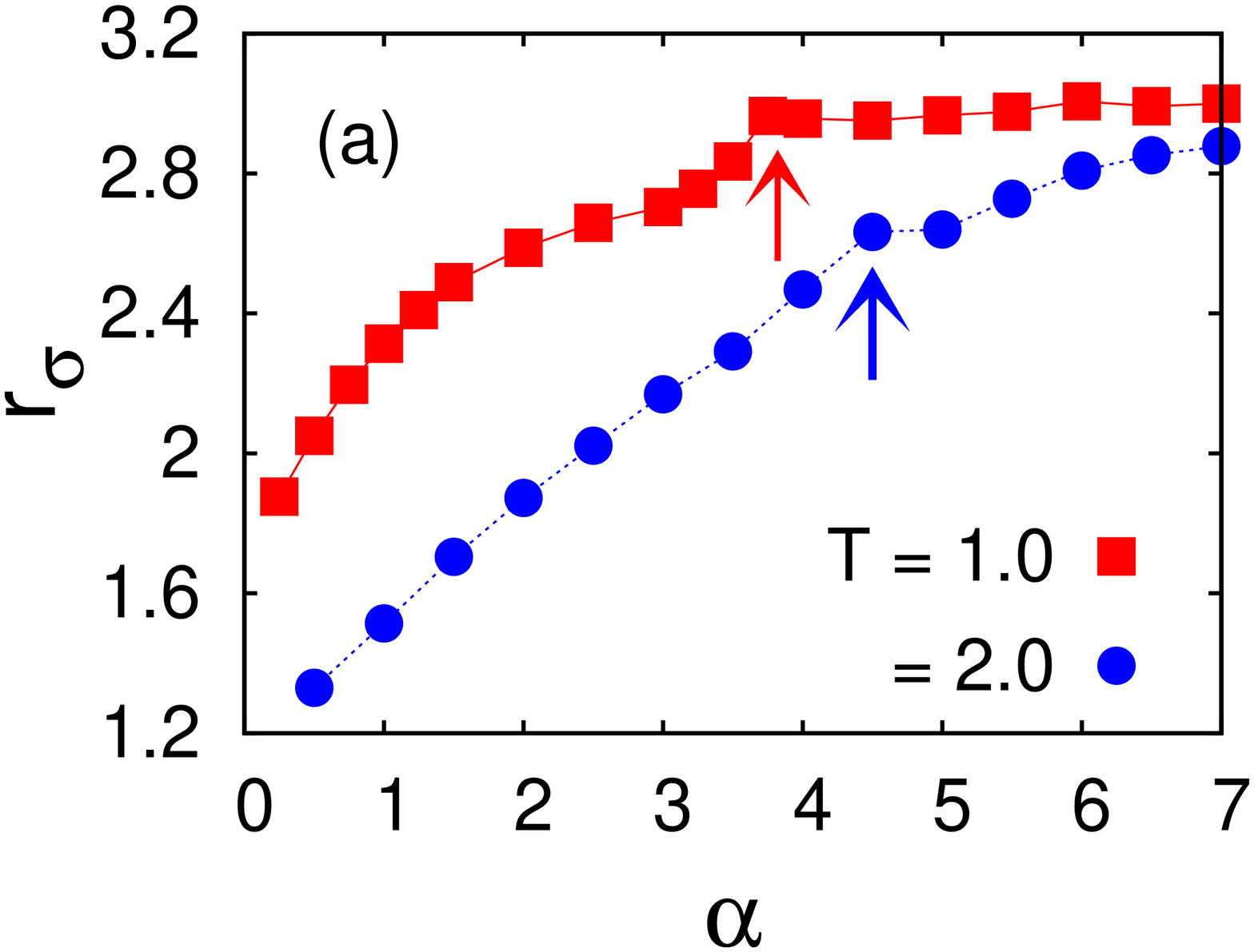}\hskip-0.3cm
\includegraphics[width=3.1cm,angle=-90]{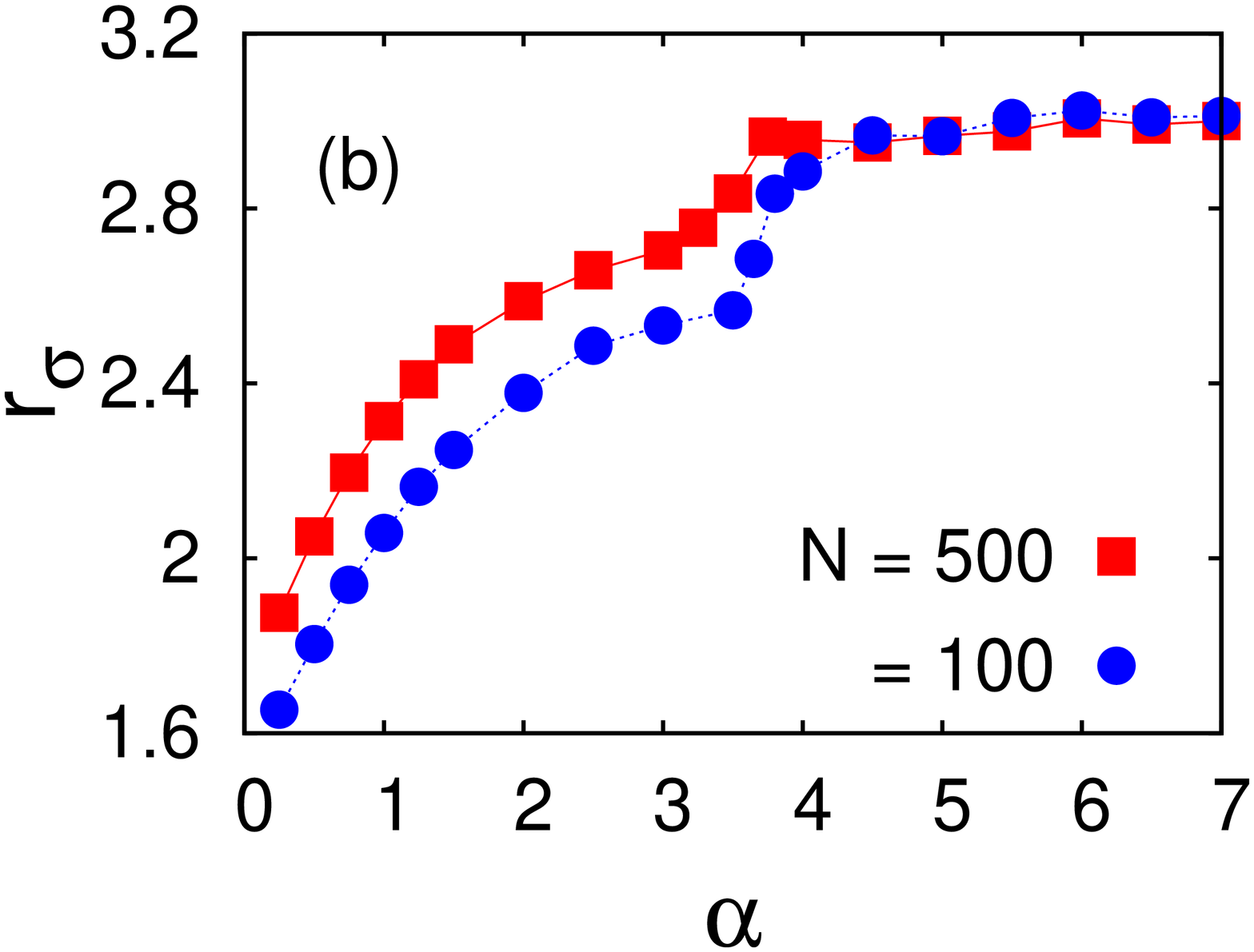}\\
\vskip-1.4cm
\includegraphics[width=5cm,angle=-90]{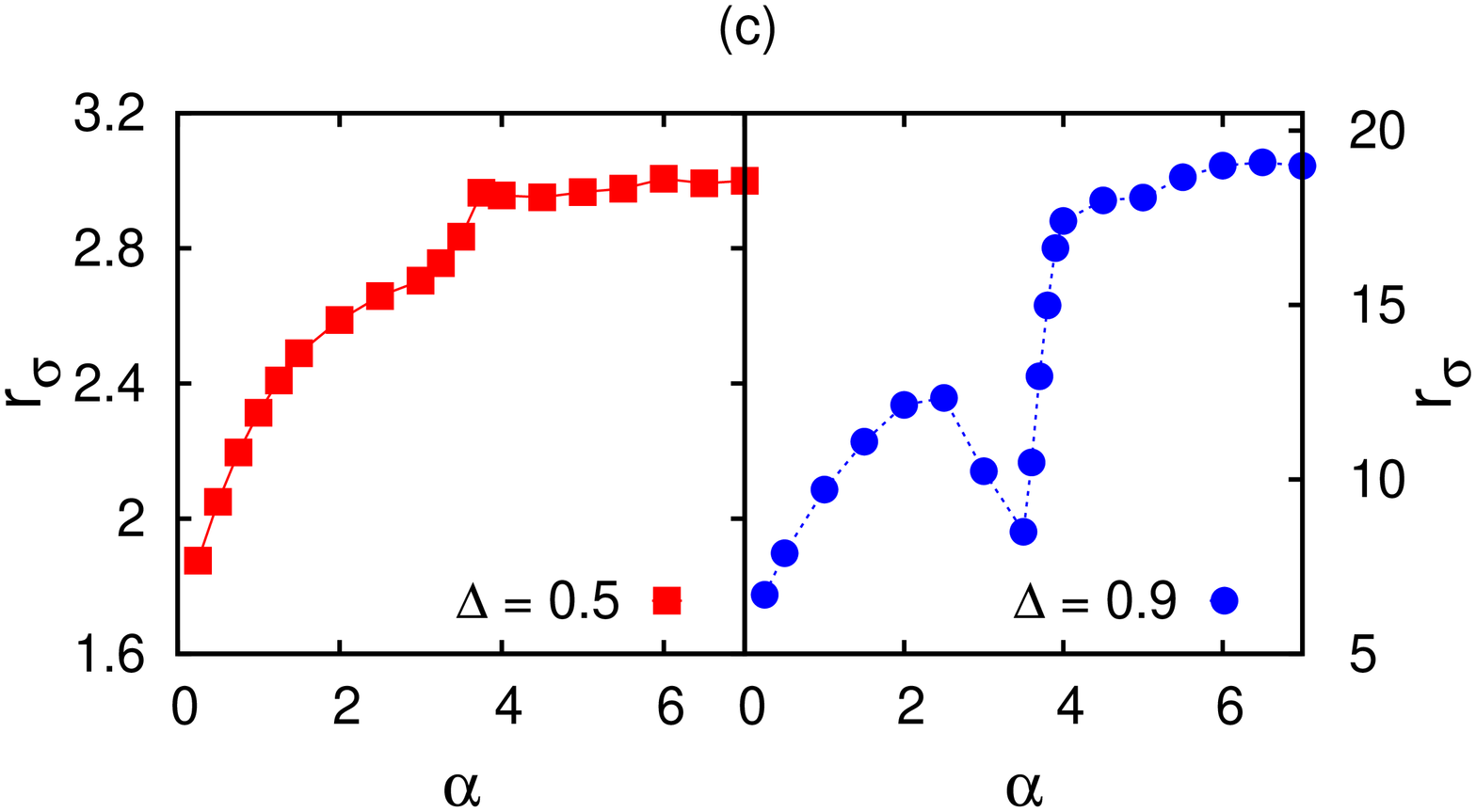}
\caption{(Color online) Variation of $r_{\sigma} = \sigma_{-\Delta}/\sigma_{\Delta}$ with $\alpha$ for 
(a) different $T = 1.0, 2.0$ with $\Delta = 0.5$ and $N = 500$; The arrows mark the onset
of the magnetic field dominated regimes.
(b) different system sizes $N = 100, 500$ - here $\Delta = 0.5$ and $T = 1.0$;
(c) different $\Delta = 0.5, 0.9$ with $T = 1.0$ and $N = 500$.
The arrows in (a) mark the onset of the magnetic field dominated regimes.}
\label{fig:sd_NDT}
\end{figure}

To obtain a numerical estimate of $\alpha_0$ where the rectification ratio $R_{\Delta} = |J_{-\Delta}/J_{\Delta}|$ shows a jump
and the current attains saturation, we compute the ratio of the two $\sigma$ values, $r_{\sigma} = \sigma_{-\Delta}/\sigma_{\Delta}$,
and study its variation as a function of $\alpha$. It is found that $r_{\sigma}$ increases steadily till some $\alpha$ value in the
range $3.5 < \alpha < 4.0$ and thereafter it attains constancy (Fig. \ref{fig:sd}d). This is indeed where the magnetic field overpowers
the thermal motion of the spins that we have encountered earlier - the individual currents attain saturation and the rectification ratio
jumps from $R_{\Delta}>1$ to $R_{\Delta}<1$.
In fact all the features that we have observed in Fig. \ref{fig:RTL}a can be understood from the $\alpha$ variation of $r_{\sigma}$.
In Fig. \ref{fig:sd_NDT} we show $r_{\sigma}$ as a function of $\alpha$ for different values of $N,T$ and $\Delta$. We find that
$\alpha_0$ 
remains the same for different system sizes $N$ and bias $\Delta$ values (Fig. \ref{fig:sd_NDT}a,c) and shifts only when the average
temperature $T$ is varied (Fig. \ref{fig:sd_NDT}b). Also note that the nonmonotonic behavior of $R_{\Delta}$ that is observed for $\alpha
< \alpha_0$ in some of the curves of Fig. \ref{fig:RTL} can be clearly seen in the plot for $r_{\sigma}$ (Fig. \ref{fig:sd_NDT}c).

Thus the physical picture that we had proposed to explain the features observed in rectification is corroborated
by the results obtained from simulation. Thermal rectification, as exhibited by this system, shows several intriguing features
that have not been observed or investigated in any of the previous works of rectification and can have a lot of technological
implications in the fabrication of thermal devices. It will also be interesting to see if such peculiar dependences arise in
other graded systems.

\subsection{Negative Differential Thermal Resistance}
Next, we turn our attention to the emergence of NDTR in this system. Note that the current $J$ in a strictly non-decreasing function
of $\Delta$ as has been obtained in  Fig. \ref{fig:rect}. To make this system exhibit NDTR, we keep the temperature of one bath fixed
and change the temperature of the other bath; we set $T_l = T$ and $T_r = T - \Delta$. The magnetic field is chosen to be linearly varying
in space $h^z_i = h_0 + \alpha ~ i/N$ as in the previous section.

\begin{figure}[h]
\hskip-0.5cm
\includegraphics[width=3.35cm,angle=-90]{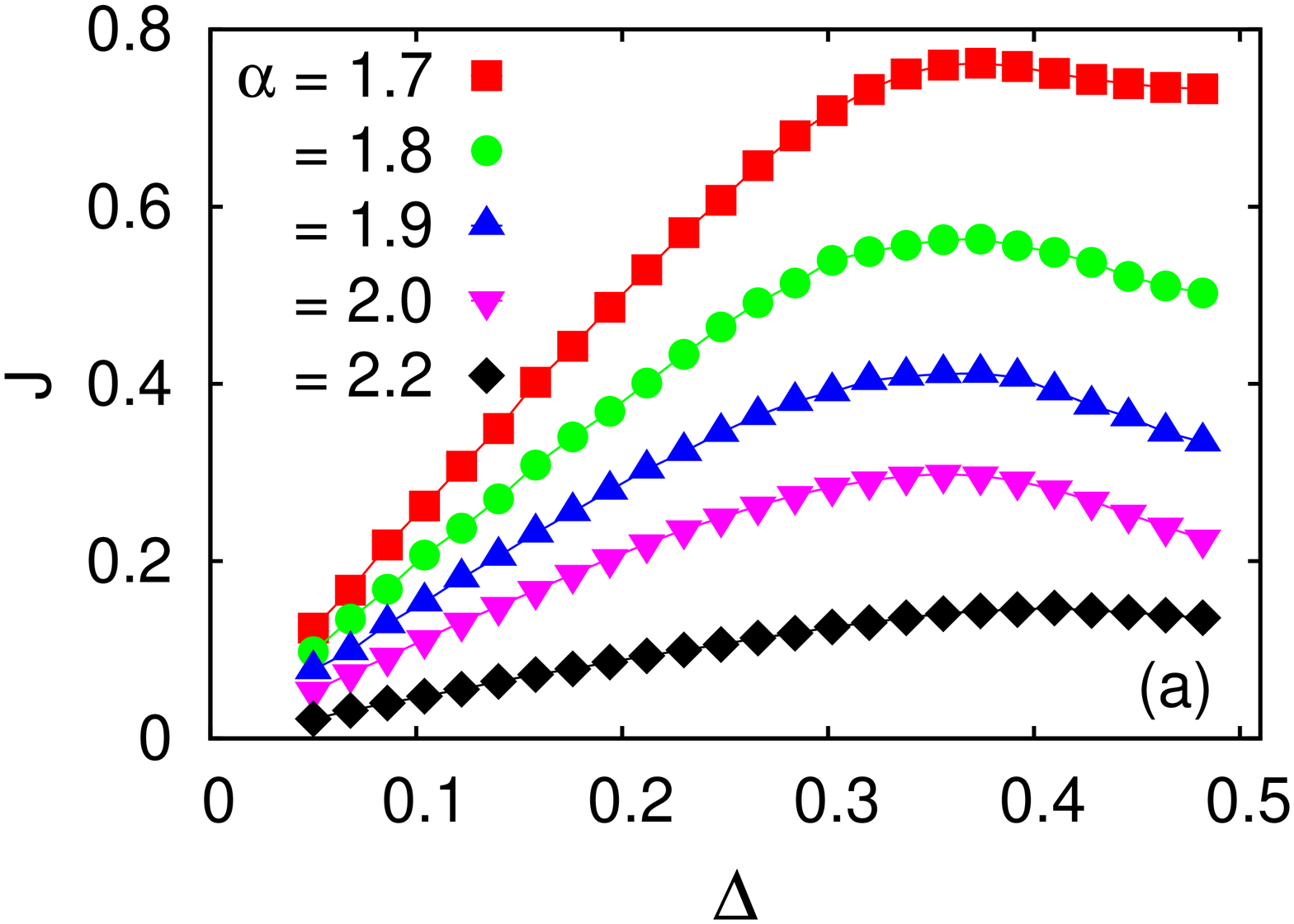}\hskip-0.5cm
\includegraphics[width=3.35cm,angle=-90]{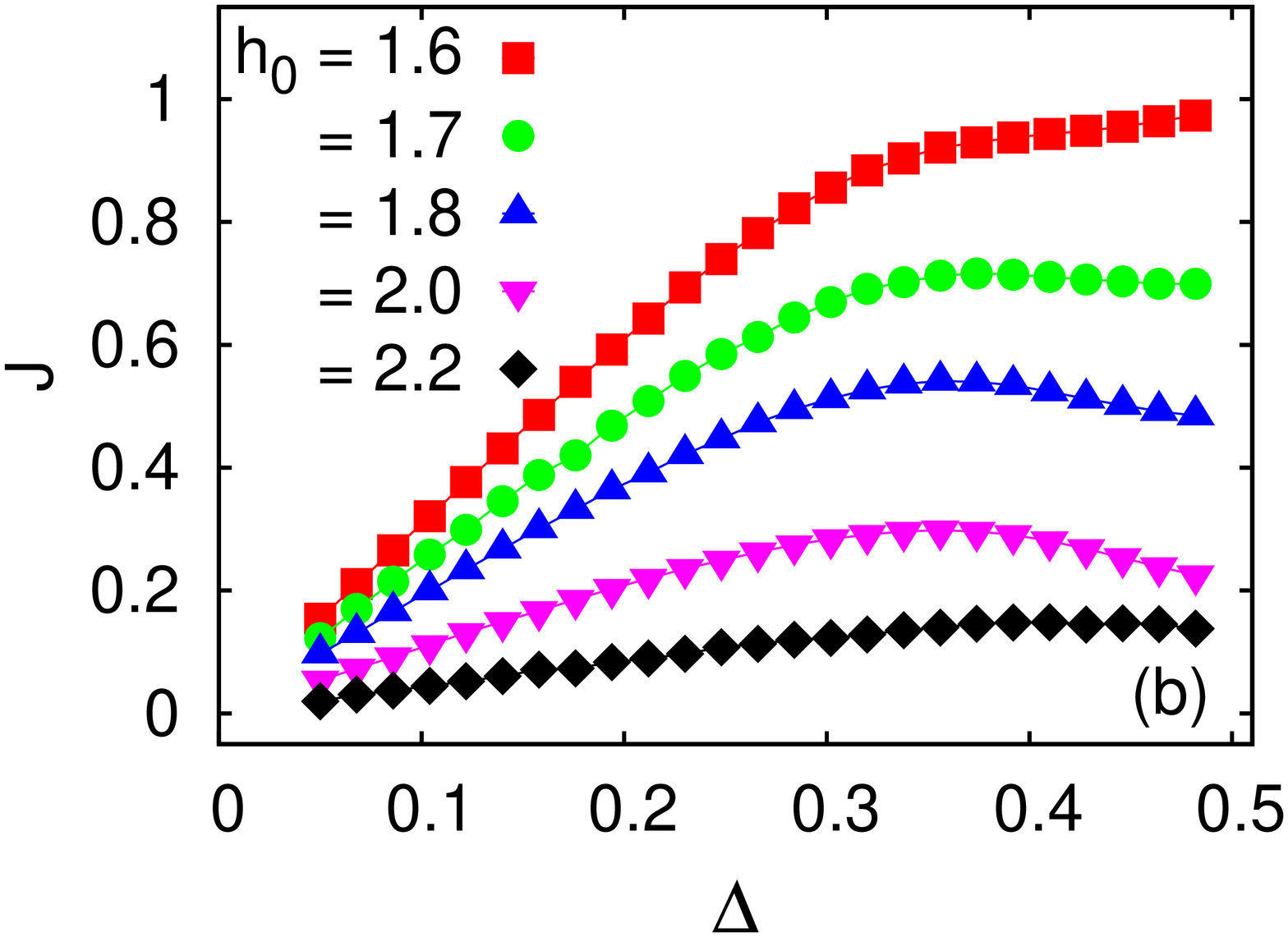}
\caption{(Color online) Variation of the total thermal current $J$ with $\Delta$ for different values of (a) $\alpha$
with $h_0 = 2$ and (b) $h_0$ with $\alpha = 2$. The temperatures are chosen as $T_l = T$ and $T_r= T-\Delta$.
For both the figures $T = 0.5$ and $N = 500$.}
\label{fig:dhh0}
\end{figure}

The variation of the total thermal current $J$ with $\Delta$ for different values of the parameter $\alpha$ is shown in Fig. \ref{fig:dhh0}a.
When $\alpha=0$ i.e., with an uniform magnetic field throughout the system, the current $J$ sharply increases as $\Delta$ is increased and there
is no NDTR (data not shown). This is due to the absence of any mechanism to restrict the passage of energy in the bulk of the system which is
required in order to observe NDTR \cite{NDTR}.

As $\alpha$ is increased, the system exhibits NDTR for some nonzero value of $\alpha$. Thus by simply tuning
the external magnetic field one obtains NDTR in this system without the need to manipulate parameters of the
system. For a fixed nonzero value of $\alpha$ one can also obtain NDTR by tuning $h_0$ as has been shown
in Fig. \ref{fig:dhh0}b. The physical mechanism that gives rise to NDTR is the obstruction to the flow of
current by the magnetic field as has been discussed in detail in a previous work \cite{NDTR}. Note that when
the magnetic field is increased (either by increasing $\alpha$ or $h_0$) further to larger values the current
becomes very small and the NDTR regime disappears.

\begin{figure}[h]
\hskip-0.425cm
\includegraphics[width=3.5cm,angle=-90]{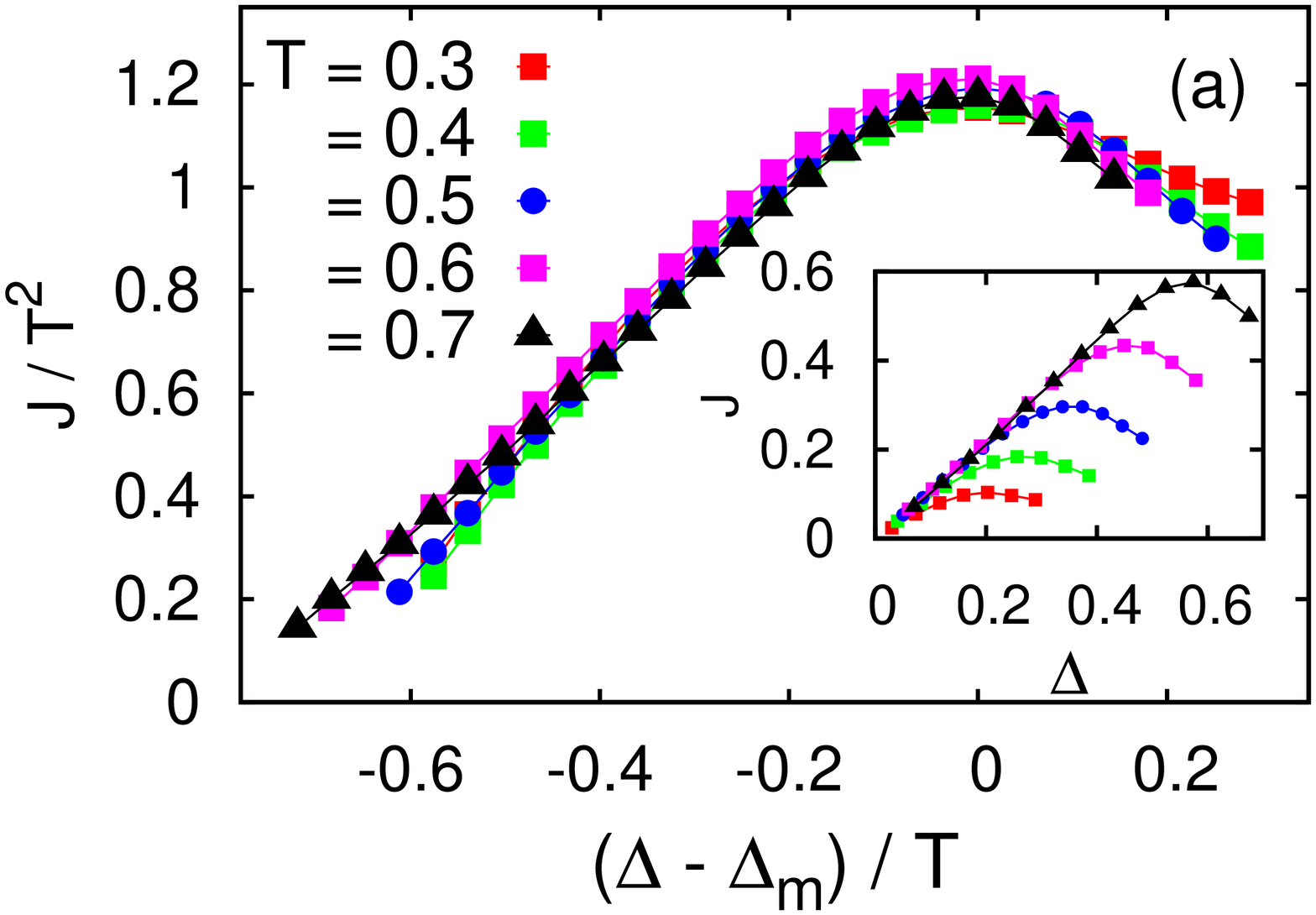}\hskip-0.53cm
\includegraphics[width=3.18cm,angle=-90]{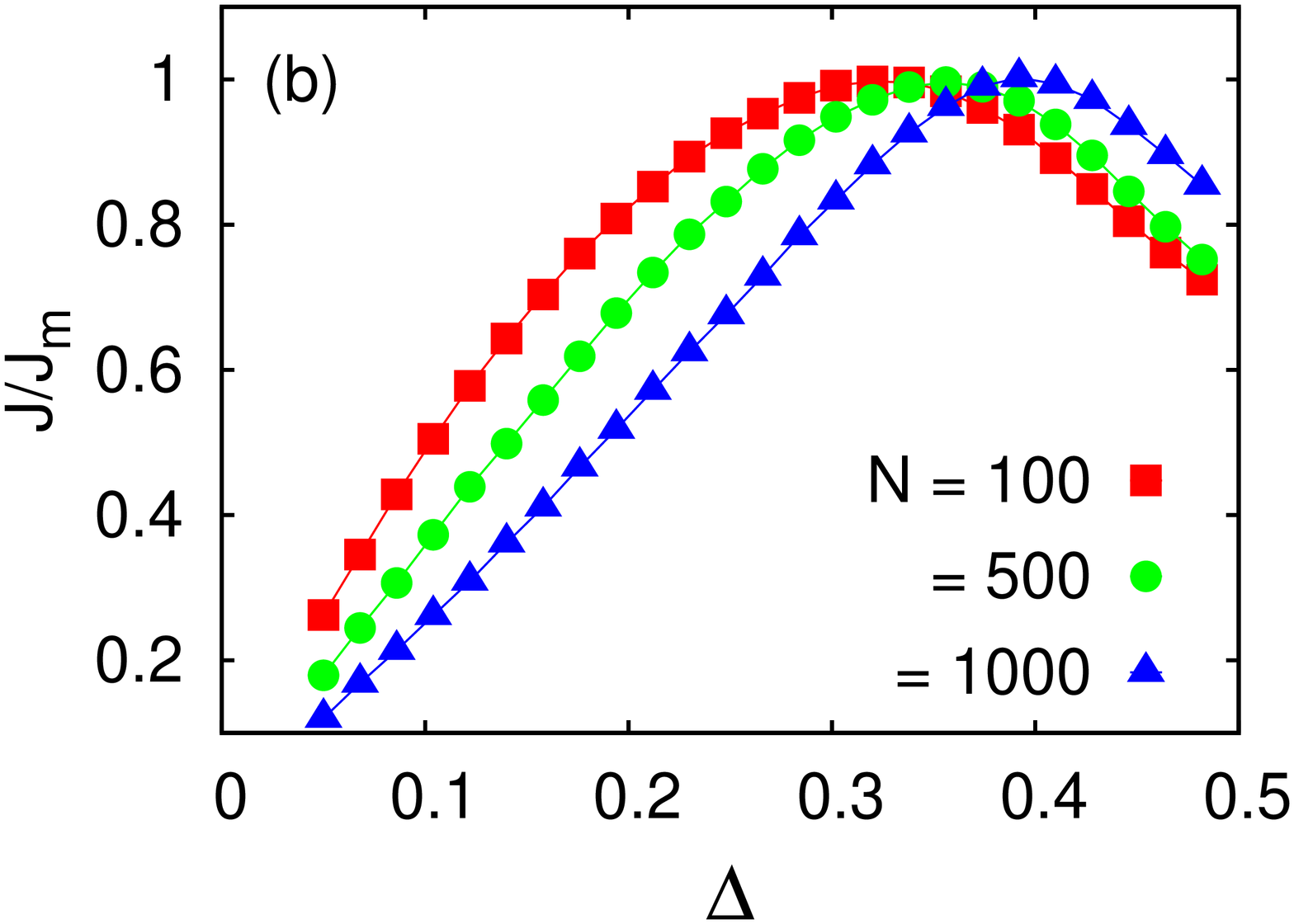}
\caption{(Color online) (a) Variation of the current $J$ with $\Delta$ for different average temperatures with system
sizes $N = 500$. (b) Variation of the current $J$ with $\Delta$ for different system sizes $N = 100, 500$
and $1000$.}
\label{fig:Tmul}
\end{figure}

The temperature dependence of NDTR is described in Fig. \ref{fig:Tmul}a. 
It is seen that the point of
emergence of NDTR $\Delta_m$ shifts to larger values of $\Delta$ as temperature increases. The
value of the energy current increases too as the temperature is increased. From the main figure we
find that the $J \sim \Delta$ curves show an excellent data collapse when the axes are rescaled as
$J/T^{\nu}$ and $(\Delta - \Delta_m)/T$; for the chosen set of parameters $\nu = 2.0$ and $\Delta_m$
is the point where NDTR regime commences corresponding to the maximum value of current.

As has been commonly seen in previous works, NDTR becomes more pronounced as the system size $N$ is
decreased. Here too, we find that the point of commencement of NDTR gradually shifts towards larger
values of $\Delta$ as the system size is increased. This is depicted in Fig. \ref{fig:Tmul}b. The
decrease of the NDTR regime due to increase in system size can however be compensated by decreasing
the temperature or increasing the magnetic field suitably.
We have also verified the emergence of NDTR in this system for other spatial dependence of the magnetic
field. With an exponentially varying magnetic field of the form $h^z_i = h_0 \exp(\alpha ~ i/N)$ we 
find a clear NDTR regime as $\alpha$ is increased from zero (data not shown). 

\section{Discussion}
\label{conclusion}
To summarise our main results, we have studied thermal rectification (TR) and negative differential thermal
resistance (NDTR) in the one dimensional classical Heisenberg model under thermal bias with a spatially varying
magnetic field. Systematic analysis of TR with respect to system parameters reveal intriguing dependences with
respect to temperature and system size. For certain range of system parameters NDTR can be observed.
Both the features emerge and can be controlled by the external magnetic field unlike the previous works where one
had to prepare the system with specific parameter values. 
%
%
%
%
%
Heat transport in magnetic system assisted by classical spin waves have been predicted several
years back \cite{Heitler} and has also been experimentally observed recently in yttrium iron
garnet \cite{Luthi, Douglass}. Transport studies in spin systems are also of active experimental
interest in recent times \cite{trans-expt1,trans-expt2}. Actual chemical compounds \cite{savin, jongh,windsor}
that mimic classical spin interactions, such as ${\rm TMMC((CH_3)_4NMnCl_3)}$ and
${\rm DMMC((CH_3)_2 NH_2 MnCl_3)}$, are already known for quite some
time now. Apart from carbon nanotubes which are considered suitable for fabricating thermal devices,
this present work (and also \cite{NDTR}) suggests these spin materials to be another promising candidate.
Hopefully, with the recent advancement in low dimensional experimental techniques, these theoretical
predictions would be verified experimentally and lead to the efficient thermal management in future.

\end{document}